\pdfoutput=1
\documentclass[conference]{IEEEtran}
\usepackage{cite}
\usepackage{graphicx, amsmath, fixmath, caption, amssymb, tabu}
\usepackage[caption=false]{subfig}
\usepackage{commath}
\usepackage{bm}
\usepackage{verbatim}

\graphicspath{ {images/} }
\ifCLASSINFOpdf
\else
\fi
\newcommand{\quotes}[1]{``#1''}

\begin{document}
%
\title{\LARGE \bf A Fuzzy Logic System to Analyze a Student's Lifestyle\\[-3.0ex]}



%
\author{\IEEEauthorblockN{Sourish Ghosh,
Aaditya Sanjay Boob,
Nishant Nikhil, 
Nayan Raju Vysyaraju, and
Ankit Kumar}
\IEEEauthorblockA{Department of Mathematics, Indian Institute of Technology Kharagpur, India\\
Email: \{\href{mailto:sourishg@iitkgp.ac.in}{sourishg}, \href{mailto:aaditya733@iitkgp.ac.in}{aaditya733}, \href{mailto:nishantnikhil@iitkgp.ac.in}{nishantnikhil}, \href{mailto:v.nayanraju@iitkgp.ac.in}{v.nayanraju}, 
\href{mailto:ankitkahnani@iitkgp.ac.in}{ankitkahnani}\}@iitkgp.ac.in}
\\}


\maketitle

\begin{abstract}
A college student\rq s life can be primarily categorized into domains such as education, health, social and other activities which may include daily chores and travelling time. Time management is crucial for every student. A self realisation of one\rq s daily time expenditure in various domains is therefore essential to maximize one\rq s effective output. This paper presents how a mobile application using Fuzzy Logic and Global Positioning System (GPS) analyzes a student\rq s lifestyle and provides recommendations and suggestions based on the results.
\end{abstract}


%
\IEEEpeerreviewmaketitle

\section{Introduction}

A college student\rq s life is multidimensional. Students are expected to be academically excellent, physically fit and socially active along with managing their daily chores and pursuing their fields of interest. This structure would not only help students to engage all activities but also help them live a balanced life. This practice would eventually help them make better career choices on the basis of their interests. For such a practice one needs to invest a threshold amount of time and effort in all the activities. However only a certain amount of students are involved and excel in such a practice. In recent times various student related issues have been addressed by researchers using fuzzy logic. Patel et al. \cite{Patel} have evaluated student\rq s academic performance considering various factors such as attendance, internal exam, lab assignments, and teamwork evaluation. Chrysafiadi and Virvou \cite{Chrysafiadi2015} have developed a fuzzy logic system which understands the forgetting process of a student. Ingoley and Bakal \cite{Ingoley} have discarded the traditional methodology of assessment of student performance by also considering personal factors such as stress and accepting the fact that the evaluating system can be non-transparent. Gokmen et al. \cite{GOKMEN2010902} have made a fuzzy evaluation system which helps to evaluate students on the basis of their performance and the type of examinations by setting up an assessment criteria before an examination. Hameed and Sorensen \cite{Hameed} have developed a reliable and robust system using Gaussian membership functions for student evaluation. Xu et al. \cite{Xu:2002:ISP:820741.820940} have personalized the web-based educational system with respect to learning materials, quiz and advices achieving effectiveness in learning. Huapaya \cite{Huapaya} has developed fuzzy student diagnosis model to help teachers evaluate students by providing a high degree of flexibility.

From the review it has been seen that not many references address the student lifestyle issues using fuzzy logic. This paper discusses a novel approach using fuzzy logic to generate an analysis of a student\rq s daily time expenditure in various categories. Based upon the analysis of the results obtained from the data appropriate recommendations and suggestions must be provided on regular basis. This would help the students work in their non-performing fields and maintain a balanced lifestyle.

\subsection{About Fuzzy Logic}

Over the past three decades, fuzzy logic is widely used in all problem-solving domains. One of the reasons for such instantaneous growth since its inception is its usability across all sectors be it Dynamic Programming, Process Control or Optimization. Fuzzy logic discards the theory of \lq Absolute Truth\rq and instead proposes a new theory of \lq Partial Truth\rq, also referred as degree of membership (suggested by Zadeh in 1965) \cite{ZADEH1965338}.

Let $\mathbf{S}$ be a non empty set, called the \textit{universe set}. Now, consider any crisp set $A \subset \mathbf{S}$. A characteristic function $\mathbold{\chi}_A$ is defined as

$$
\mathbold{\chi}_A(x) = 
\begin{cases}
    1, & \text{if } x\in A\\
    0, & \text{otherwise}
\end{cases}
$$

A characteristic function assigns value of either $0$ or $1$ to each element of $\mathbf{S}$. Now consider a fuzzy set $B \subset \mathbf{S}$. A membership function $\mathbold{\mu}_B(x)$ is defined as $\mathbold{\mu}_B:\mathbf{S}\rightarrow[0, 1]$. Unlike the notion of a set in classical set theory where an element either belongs or does not belong to a particular set based on a bivalent condition, in fuzzy set\cite{Klir:1996:FSF:234347} theory an element\rq s belongingness to a particular set is decided using membership function which gives a membership value between $0$ and $1$.

\subsection{Problem Formulation}

The problem can be divided into three major parts:

\begin{itemize}
\item \textbf{Data Collection}: Using GPS and Google Places API, data collection of all the locations visited and time spent at each location by the user.
\item \textbf{Fuzzification}: Fuzzify the crisp input and calculate the values of corresponding membership functions.
\item \textbf{Defuzzification}: Set up a fuzzy inference system based on certain rules and then return recommendations and suggestions.
\end{itemize}

\begin{figure}[h!]
\centering
\captionsetup{justification=centering}
\noindent \includegraphics[scale=0.45]{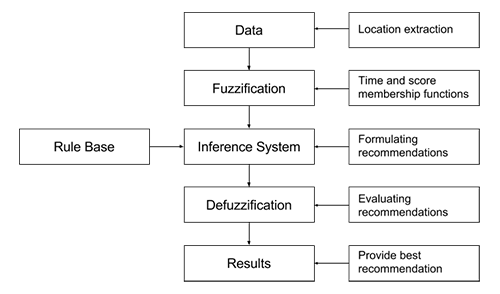}
\caption{Fuzzy logic system architecture for determining a student\rq s lifestyle.}
\end{figure}

\section{Working Principle}

The general work flow of this paper is shown in Fig. 1.

\subsection{Data Collection}

A college student is carrying his/her smart phone almost everywhere. Hence using the GPS extraction of his/her position throughout the day is possible. In the application and testing of this paper, the mobile application was developed on Android while the point of interest was extracted using Google Maps API by querying the user\rq s location extracted from GPS. Google Maps API classifies most of the locations into various categories namely \texttt{restaurant}, \texttt{shopping\_mall}, \texttt{city\_hall} etc. Let us refer to all these categories hence forward as \textit{tags}. Apart from these existing tags, two additional tags namely \texttt{home} and \texttt{work} are generated. The GPS data for these two additional tags would be user specific. Hence initially every user needs to update their location for these two tags specifically. This step is conducted so as to recognize distinctly one\rq s home and workplace which in further course would generate accurate results. Consider this example where one might go to a pizza shop to hang out with friends and family. However if someone is working in a pizza shop and the GPS details of the specific pizza shop is not known beforehand, it is very likely that one might consider this entire working time as time utilized for hanging out with friends. However if the person goes to some other pizza shop it is very likely he is going out with friends. To avoid this confusion this initial step has to be carried out.

Let $\mathbf{X}$ be the set of all tags defined as $\mathbf{X} = \{x \mid x\text{ is a tag}\}$. Analysing the way a person lives is governed by many parameters, but in a typical student\rq s life we are mainly concerned about one\rq s health, education, leisure and social life. However a person also invests certain amount of time which fails to fall under these categories. A example of this would be travelling time. Activities like these fall under the \textit{other} category. Now let $S$, $L$, $H$, $W$, $O$ be subsets of $\mathbf{X}$ defined as
\begin{align*}
S =\ &\{x \mid x \in \mathbf{X},\ x = \text{social}\ \textrm{and}\ x \neq \text{home, work}\} \\
L =\ &\{x \mid x \in \mathbf{X},\ x = \text{leisure}\ \textrm{and}\ x \neq \text{home, work}\} \\
H =\ &\{x \mid x \in \mathbf{X},\ x = \text{health}\ \textrm{and}\ x \neq \text{home, work}\} \\
W =\ &\{x \mid x \in \mathbf{X},\ x = \text{work}\ \textrm{and}\ x \neq \text{home}\} \\
O =\ & \{x \mid x \in \mathbf{X}\ \textrm{and}\ x \notin S\cup L\cup H\cup W \}
\end{align*}
A \textit{tag} $x$ might belong to one or more of the sets $S,\ L,\ H,\ W$. For example, a person might visit an Amusement Park. In this case the person\rq s \textit{social} and \textit{leisure} purposes are fulfilled. Using this categorization technique we can extract one\rq s location and time spent at each \textit{tag} for the entire day. TABLE I lists down some locations and their possible purpose of visits. The locations mentioned are basically \textit{tags} other than \texttt{home} and \texttt{work}.
\begin{table}
\small
\captionof{table}{Sample locations and purposes}
\begin{center}
\def\arraystretch{1.7}
\begin{tabular}{|l|l|}
\hline
 \bf Location & \bf Purpose \\
 \hline
 Cafe, Restaurant & Going out with friends and family. \\
 \hline
 Supermarket, Gas Station & Chores \\
 \hline
 Gym, Ground, Hospital & Exercise or Health Treatment \\
 \hline
 Cinema Hall, Spa & Leisure and Relaxation \\
\hline
Bank, Business Associates & Work \\
\hline
\end{tabular}
\end{center}
\end{table}

\textbf{Weighing criteria}: For a given purpose, different locations would have different amount of productivity and impact. For example, hospital and gym both fall under the \textit{health} category. However one visits a gym to increase his physical activity and hence visiting a gym has a positive impact on one\rq s health. However one visits a hospital if he/she has fallen sick. Hence, visiting a hospital has a negative impact on one\rq s health. So we have to handle these two situations differently.

A function $Y$ is defined as $Y:\mathbf{X} \rightarrow \mathbb{R}$ such that $Y(x)$ for every $x \in \mathbf{X}$ denotes the time spent at the location \textit{tag} $x$. For example, let $x = \text{gym}$. Say $Y(x) = 0.5$. This implies a person has spent $30$ minutes at a \textit{gym} in the entire day. The unit of time is set in hours throughout this paper.

A function $Z_S$ is defined as $Z_S:S \rightarrow [-100, 100]$ such that $Z_S(x)$ for every $x \in S$ denotes the intensity of the \textit{tag} $x$ with respect to the \textit{social} category. Similarly one can define $Z_H$, $Z_L$, $Z_W$, and $Z_O$ for the \textit{health}, \textit{leisure}, \textit{work}, and \textit{other} categories respectively. The range $[-100, 100]$ is chosen for normalization purposes. For example, let $x = \text{gym}$. Say $Z_H(x) = 50 > 0$ as gym has a positive health impact. Let $y = \text{hospital}$, then $Z_H(y) = -20 < 0$ as hospital has a negative health impact. However, $Z_L(x) = Z_L(y) = 0$ as both $x$ and $y$ don\rq t contribute to the \textit{leisure} category. Also note that if a \textit{tag} $t$ belongs to two different categories, then its weightage in both the categories cannot be $0$.

For both $Y$ and $Z$ don\rq t include the \textit{home} tag as it is a special case. This is explained later.

\begin{figure}[h!]
\centering
\captionsetup{justification=centering}
\noindent \includegraphics[scale=0.5]{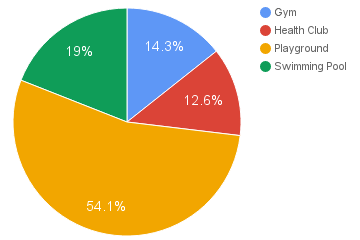}
\caption{Survey for \textit{positive} health weights}
\end{figure}
\begin{figure}[h!]
\centering
\captionsetup{justification=centering}
\noindent \includegraphics[scale=0.5]{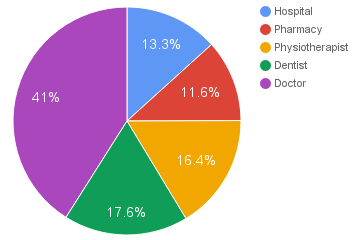}
\caption{Survey for \textit{negative} health weights}
\end{figure}
\textbf{Assigning weights}: One is free to assign the weights independently. However for better results, one can assign weights by conducting a survey to understand how appropriately a location \textit{tag} fulfils the purpose of a category. For instance consider the \textit{health} category. In the survey a sample population was asked to rank every $x \in H$ in an order of fulfilment of their positive \textit{health} benefits. Consider the following survey with 
\begin{align*}
H =\ \{&\text{gym},\ \text{playground},\ \text{swimming\_pool}, \text{health\_club},\\
&\text{hospital},\ \text{pharmacy},\ \text{physiotherapist},\ \text{dentist},\ \text{doctor}\}
\end{align*}

Fig. 2 shows a survey for determining \textit{positive} weights in the \textit{health} category. As $54.1\%$ people taking the survey voted \textit{playground} as their maximum positive benefit from the \textit{health} category, the corresponding weight for $x = \text{playground}$ is computed as $Z_H(x) = \frac{54.1}{100} \times 100 = 54.1$.

\begin{figure*}
\subfloat[Membership function of \textit{social} category]{%
  \includegraphics[scale=0.4]{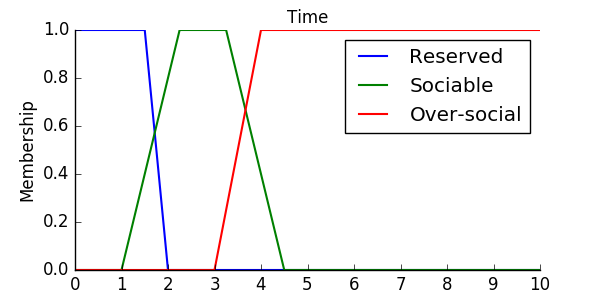}%
}\hfill
\subfloat[Membership function of \textit{leisure} category]{%
  \includegraphics[scale=0.4]{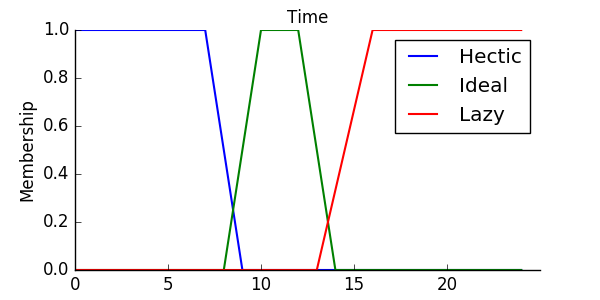}%
}\hfill
\subfloat[Membership function of \textit{other} category]{%
  \includegraphics[scale=0.4]{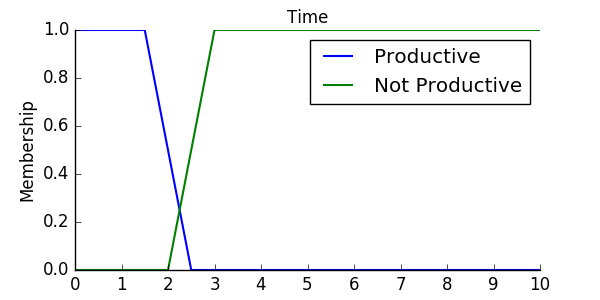}%
}\hfill
\subfloat[Membership function of \textit{work} category]{%
  \includegraphics[scale=0.4]{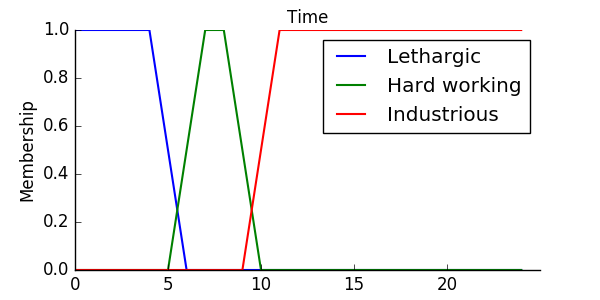}%
}\hfill
\subfloat[Membership function of \textit{health} category]{%
  \includegraphics[scale=0.4]{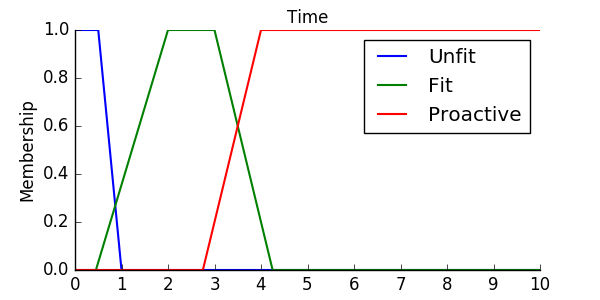}%
}\hfill
\subfloat[Membership function of the linguistic term \textit{fit}]{%
  \includegraphics[scale=0.4]{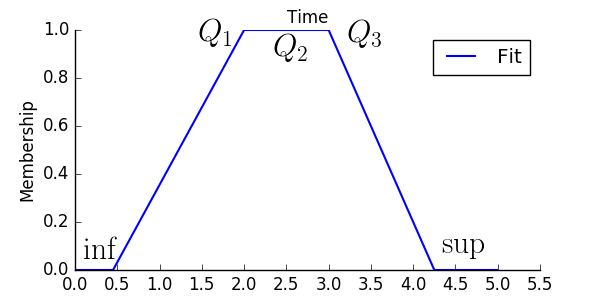}%
}
\caption{Membership functions for all linguistic terms of all categories with respect to time.}
\end{figure*}

Fig. 3 shows a survey for determining \textit{negative} weights in the \textit{health} category. As $41\%$ people taking the survey voted \textit{doctor} as their maximum non fulfilment from the \textit{health} category, the corresponding weight for $x = \text{doctor}$ is computed as $Z_H(x) = -\frac{41}{100} \times 100 = -41$.

\textbf{Home tag}: The time spent at the \textit{home} location might not be entirely used for rest and leisure purpose only. One might practice yoga at one\rq s home and the equivalent time should be added to the \textit{health} category. Let $\tau$ denote the total time spent at \textit{home}. And $\tau_H$, $\tau_W$, $\tau_L$, $\tau_O$, $\tau_S$ denote the equivalent time in respective categories. This time is taken as user input through the mobile application. For better results a random push notification system is used to learn the characteristics of the user. The home tag will be associated with weights $\xi_H$, $\xi_S$, $\xi_L$, $\xi_W$, $\xi_O$ which denote the intensity of the tags at \textit{home}. For instance, $\xi_W=30$ and $Z_W(\text{office}) = 50 > 30$ as working at \textit{home} might not be as productive as working at \textit{office}.

\subsection{Fuzzification}

\textbf{Fuzzification of time}: Consider a person $p$. Suppose $p$ visits tags $\{x_1, x_2, \ldots, x_n\}$, with the time spent at these locations denoted by $\{Y(x_1), Y(x_2), \ldots, Y(x_n)\}$. Let $K_H$, $K_L$, $K_S$, $K_W$, $K_O$ denote the overall time spent in \textit{health}, \textit{leisure}, \textit{social}, \textit{work}, and \textit{other} categories respectively. Then 
\[
K_H = \sum_{x_i \in H}Y(x_i) + \tau_H
\]
Similarly, $K_W$, $K_S$, $K_L$, $K_O$ are defined.

We define the following fuzzy sets for all the categories. These sets define the type of lifestyle a person is living in each category. Here \textit{leisure} also includes rest.
\begin{align*}
\textit{health}=&\ \{\text{unfit},\ \text{fit},\ \text{proactive}\}\\
\textit{leisure}=&\ \{\text{hectic},\ \text{ideal},\ \text{lazy}\}\\
\textit{social}=&\ \{\text{reserved},\ \text{sociable},\ \text{over\_social}\}\\
\textit{work}=&\ \{\text{lethargic},\ \text{hard\_working},\ \text{industrious}\}\\
\textit{others}=&\ \{\text{non\_productive},\ \text{productive}\}
\end{align*}
The membership functions for these fuzzy sets are constructed by conducting a survey on a sample population. The data from the survey can be approximated by using quantile range and trapezoidal membership functions. However, one can use various other techniques to plot membership functions. For instance, in a sample survey the hours spent by fit students in the \textit{health} category were: $0.45,\ 1.25,\ 2,\ 2.25,\ 2.5,\ 2.5,\ 2.75,\ 2.75,\ 3,\ 4,\ 4.25$. So with respect to the inter quantile range $Q_1 = 2$, $Q_2 = 2.5$, $Q_3 = 3$, $\inf = 0.45$, and $\sup = 4.25$. The trapezoidal membership function for the linguistic term \quotes{fit} under the \textit{health} category using these values is shown in Fig. $4$.

Fig. $4$ shows the membership functions for each linguistic of all categories.

\textbf{Fuzzification of score}: Not only the time spent at a location is important but also how the time is spent is important too. This effective utilisation of time is denoted by a \textit{score} $M_S$, $M_L$, $M_O$, $M_W$, and $M_H$ for the respective categories. The score for the \textit{social} category is calculated as follows
\[
M_S = \sum_{x \in S} Y(x)Z_S(x) + \tau_S\xi_S
\]
Similarly other scores can be defined. The fuzzy set of linguistic terms \quotes{low\_score}, \quotes{ideal\_score} and \quotes{high\_score} define the fuzzy scores in each category. The membership function of these sets in all categories is calculated similar to the fuzzy time membership functions by conducting a survey. For instance, a survey conducted on a sample of fit students is shown in TABLE II.
\begin{table}
\small
\captionof{table}{Survey to determine the membership function for the linguistic term \quotes{ideal\_score} under \textit{health} category}
\begin{center}
\def\arraystretch{1.7}
\begin{tabular}{| l | l | l |}
\hline
\bf Time $(t_i)$ & \bf Weight $(w_i)$ & \bf Score $(\sum_{i} w_it_i)$ \\
\hline
$0.45$ & $25$ & $0.45\times 25=11.25$ \\
\hline
$1$, $0.25$ & $10,\ 12$ & $1\times10+0.25\times 12=13$ \\
\hline
$2$ & $15$ & $2\times 15=30$ \\
\hline
$2.25$ & $18$ & $2.25\times 18=40.5$ \\
\hline
$2.5$ & $18$ & $2.5\times 18=45$ \\
\hline
$2.5$ & $20$ & $2.5\times 20=50$ \\
\hline
$2.75$ & $12$ & $2.75\times 12=33$ \\
\hline
$2,\ 0.75$ & $13,\ 10$ & $2\times 13+0.75\times 10=33.5$ \\
\hline
$3$ & $11$ & $3\times 11=33$ \\
\hline
$2,\ 2$ & $13,\ 8$ & $2\times 13+2\times 8=42$ \\
\hline
$4.25$ & $7$ & $4.25\times 7 =29.75$ \\
\hline
\end{tabular}
\end{center}
\end{table}
Hence $\inf = 11.25,\ Q_1=29.75,\ Q_3=42,\ \sup = 50$.  Accordingly the membership function for the linguistic term \quotes{ideal\_score} under the health category is shown in Fig. 5. Similarly one can plot plot the membership functions for the entire fuzzy set across all categories.
\begin{figure}[h!]
\centering
\captionsetup{justification=centering}
\noindent \includegraphics[scale=0.5]{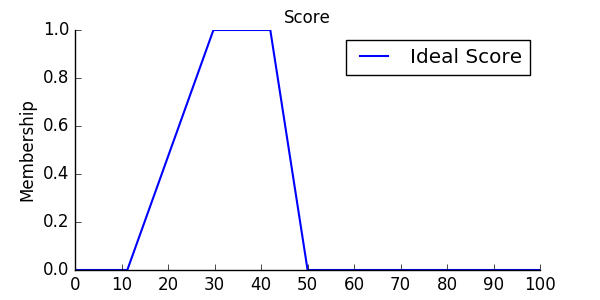}
\caption{Membership function for \quotes{ideal\_score} under \textit{health} category}
\noindent \includegraphics[scale=0.5]{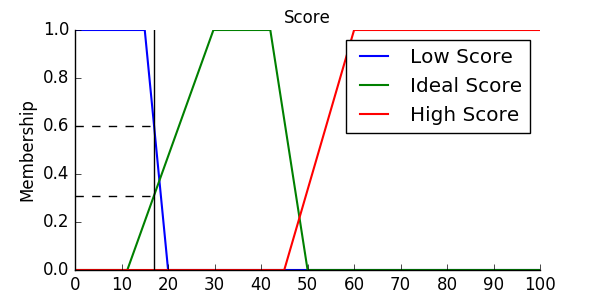}
\caption{Calculation of membership values}
\end{figure}
\subsection{Defuzzification}
Given the input data, $Y(x)$, $Z_i(x)$, $\tau_i$, and $\xi_i$ where $x \in \mathbf{X}$ and $i = S,\ L,\ O,\ W,\ H$, one calculate corresponding $K_i$ and $M_i$. Using surveys the membership functions for all linguistic terms in all categories for both fuzzification of time and score can be determined. Hence the membership value of $K_i$ and $M_i$ in the respective categories for all the linguistic terms can be determined. Let $\{R_1, R_2, \ldots, R_N\}$ be a set of recommendations. Now every $R_k\ (1 \leq k \leq N)$ will be dependent on a set of linguistic terms. For example, a recommendation $R = \textit{\quotes{All work and no play makes Jack a dull boy.}}$ will be outputted if a person is spending too much time and effort in work and less in his leisure and social life. That is he/she has a \quotes{industrious} work life with a high work score and has a \quotes{reserved} social life with a low social score and a \quotes{hectic} life with respect to leisure with a low score. So attributes of $R$ can be represented as $\{K_W=\text{\quotes{industrous}},\ M_W=\text{\quotes{high\_score}},\ K_S=\text{\quotes{reserved}},\ M_S=\text{\quotes{low\_score}},\ K_L=\text{\quotes{hectic}},\ M_L=\text{\quotes{low\_score}}\}$.

Let $R_k$ be a recommendation with attributes $\{a_1, a_2, a_3, \ldots, a_n\}$. Here each $a_j\ (1 \leq j \leq n)$ is a combination of score/time with respect to a linguistic term of a category. Hence as shown previously one can calculate its membership value. Let $\mu_1, \mu_2, \mu_3, \ldots, \mu_n$ denote the respective membership values for each attribute. Here $n$ can vary for each $R_k$.  For instance, Fig. 6 shows the membership functions of $M_H$. Let $a_1$, $a_2$, $a_3$ be the following attributes
\begin{align*}
a_1\ =&\ M_H:low\_score\\
a_2\ =&\ M_H:ideal\_score\\
a_3\ =&\ M_H:high\_score
\end{align*}
Hence $\mu_1$, $\mu_2$, $\mu_3$ for $M_H=17$ as seen from Fig. 6 will be $0.6$, $0.310$, $0.0$ respectively. Using equal weighing criteria for each $a_j$, we can calculate a score of each recommendation $\rho(R_k)$ defined as
$$\rho(R_k) = \frac{1}{n}\sum_{j=1}^{n} \mu_j$$
Now, using the \textit{most probable criterion} the recommendation with the maximum score value $\rho(R_k)$ will be displayed as output.

\section{Experiment}
A survey conducted in IIT Kharagpur was conducted to determine all the membership functions for all linguistic terms across all the categories. Some of the membership functions are shown in this paper. The mobile application was installed on the student\rq s phone and the results were analysed. A random student was picked and his data for the day was analysed. TABLE III shows the \textit{tags} he visited throughout the day and their corresponding time and weights. The score for each \textit{tag} is also enumerated. TABLE IV shows the total time and score across all the categories. The recommendations in the set $R$ where $R = \{R_1, R_2, R_3, R_4\}$ were considered.
\begin{align*}
R_1 =\ &\textit{\quotes{Catch up a movie this evening.}}\\
R_2 =\ &\textit{\quotes{Work is worship.}}\\
R_3 =\ &\textit{\quotes{Family matters.}}\\
R_4 =\ &\textit{\quotes{Hit the gym.}}
\end{align*}
The attributes of $R$ is shown in TABLE V. The membership values for each attribute is shown in TABLE VI and the corresponding score of each recommendation is also enumerated. As $\rho(R_1)$ is maximum the mobile application recommended the student to \textit{\quotes{Catch up a movie this evening.}}

The mobile application was developed on Android platform using Google Maps API for location tracking via GPS. The screenshots of the application developed are shown in Fig. 7.
\begin{figure}[!tbp]
  \centering
  \subfloat[Recommendation screen]{\includegraphics[width=4cm]{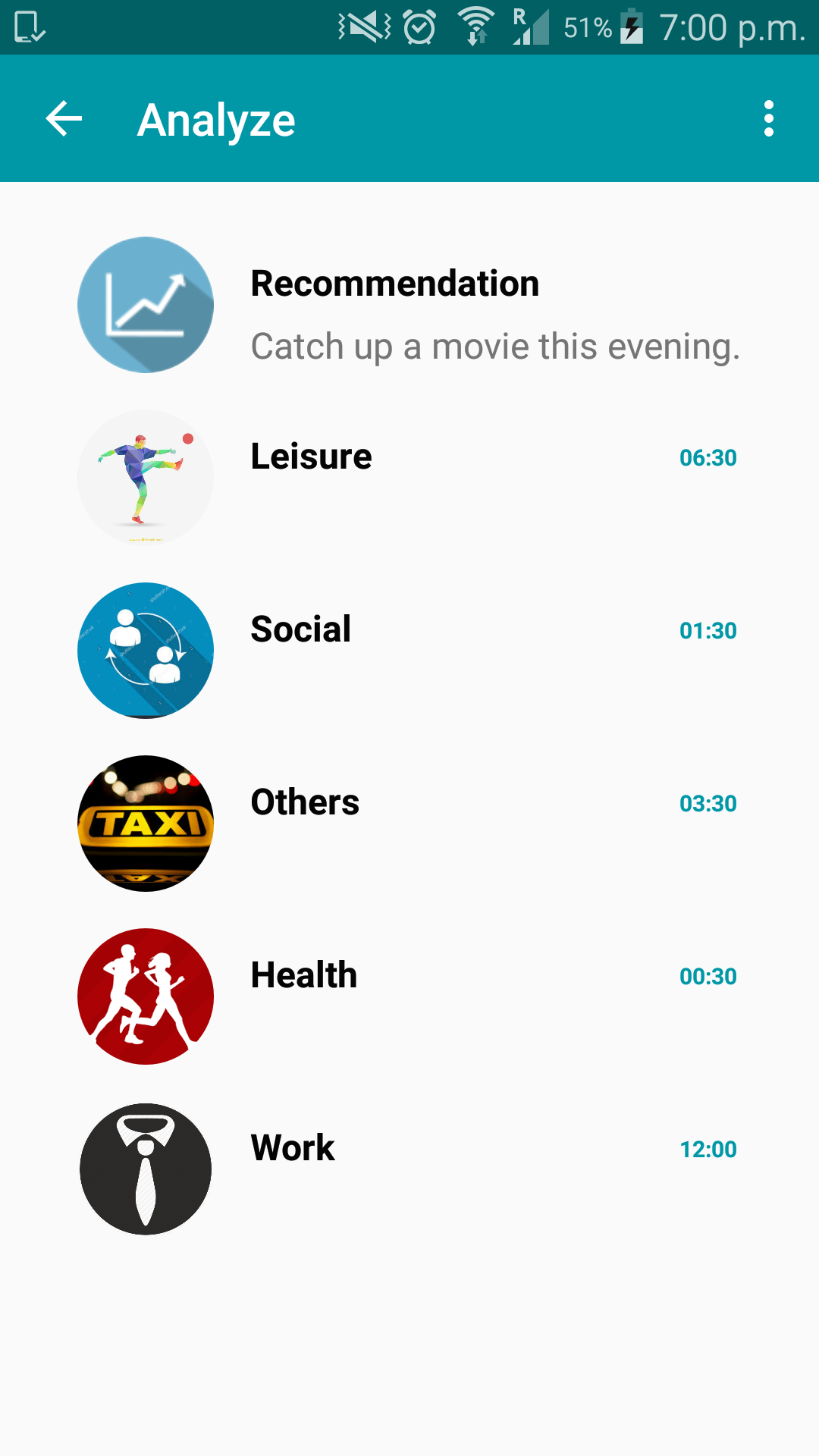}}
  \hfill
  \subfloat[\textit{Work} category time distribution ]{\includegraphics[width=4cm]{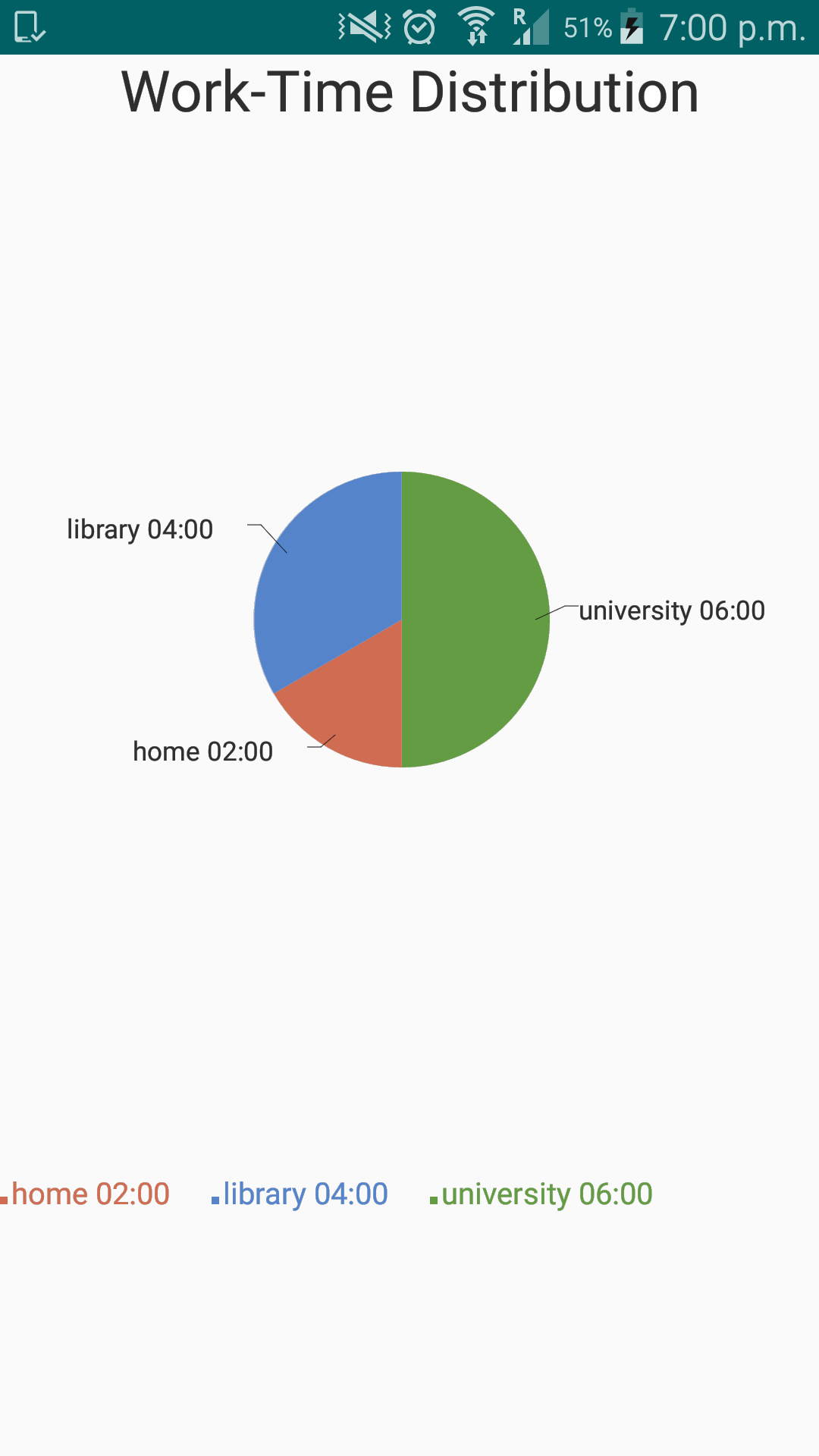}}
  \caption{Screenshots of the mobile application}
\end{figure}
\begin{table}
\small
\captionof{table}{Experiment: Data}
\begin{center}
\def\arraystretch{1.7}
\begin{tabular}{| l | l | l | l |}
\hline
\bf Tag & \bf Time & \bf Weight & \bf Score \\
\hline
university & $Y(x)=6$ & $Z_W(x) = 50$ & $300$ \\
\hline
library & $Y(x)=4$ & $Z_W(x) = 20$ & $80$ \\
\hline
home & $\tau_W=2$ & $\xi_W=30$ & $60$ \\
& $\tau_S=0.5$ & $\xi_S=30$ & $15$ \\
& $\tau_H=0.5$ & $\xi_H=20$ & $10$ \\
& $\tau_L=6.5$ & $\xi_S=30$ & $195$ \\
& $\tau_O=1$ & $\xi_O=10$ & $10$ \\
\hline
cafe & $Y(x)=1$ & $Z_S(x)=20$ & $20$ \\
\hline
supermarket & $Y(x)=1$ & $Z_O(x)=9$ & $9$ \\
\hline
grocery & $Y(x)=0.5$ & $Z_O(x)=10$ & $5$ \\
\hline
travel & $Y(x)=1$ & $Z_O(x)=15$ & $15$ \\
\hline
\end{tabular}
\end{center}
\vspace{-1em}
\end{table}
\begin{table}
\small
\captionof{table}{Experiment: Calculation of total time and score for each category}
\begin{center}
\def\arraystretch{1.7}
\begin{tabular}{| l | l |}
\hline
\bf Total time & \bf Total score \\
\hline
$K_S=1.5$ & $M_S=35$ \\
\hline
$K_L=6.5$ & $M_L=195$ \\
\hline
$K_O=3.5$ & $M_O=39$ \\
\hline
$K_W=12$ & $M_W=440$ \\
\hline
$K_H=0.5$ & $M_H=10$ \\
\hline
\end{tabular}
\end{center}
\vspace{-1em}
\end{table}
\begin{table}
\small
\captionof{table}{Experiment: Recommendation attributes}
\begin{center}
\def\arraystretch{1.7}
\begin{tabular}{| l | l |}
\hline
\bf Recommendation & \bf Attributes \\
\hline
$R_1$ & $\{K_L = \text{\quotes{hectic}},\ M_L = \text{\quotes{less\_score}},$\\&$\ K_W=\text{\quotes{industrious}},\ M_W=\text{\quotes{high\_score}}\}$ \\
\hline
$R_2$ & $\{K_W = \text{\quotes{lethargic}},\ M_W = \text{\quotes{less\_score}},$\\& $\ K_L=\text{\quotes{lazy}},\ M_L=\text{\quotes{high\_score}}\}$ \\
\hline
$R_3$ & $\{K_S = \text{\quotes{reserved}},\ M_S = \text{\quotes{less\_score}}\}$ \\
\hline
$R_4$ & $\{K_H = \text{\quotes{unfit}},\ M_H = \text{\quotes{less\_score}}\}$ \\
\hline
\end{tabular}
\end{center}
\vspace{-1em}
\end{table}
\begin{table}
\small
\captionof{table}{Experiment: Recommendation score calculation}
\begin{center}
\def\arraystretch{1.7}
\begin{tabular}{| l | l | l |}
\hline
\bf Recommendation & \bf Membership Values $(\mu_{j})$ & \bf Score $(\rho(R_k))$ \\
\hline
$R_1$ & $\{1.0,\ 0.8,\ 1.0,\ 1.0\}$ & $0.95$ \\
\hline
$R_2$ & $\{0.0,\ 0.0,\ 0.0,\ 0.0\}$ & $0.0$ \\
\hline
$R_3$ & $\{1.0,\ 0.7\}$ & $0.85$ \\
\hline
$R_4$ & $\{1.0,\ 0.8\}$ & $0.9$ \\
\hline
\end{tabular}
\end{center}
\end{table}


%
\section{Concluding Remarks}
Time management has always been a difficult art to master. This paper helps one master it by using fuzzy logic to understand the science behind this art. The use of this technology in the long run would lead to more accurate results. The main highlights of this method is that most of the segments are self configurable. The dynamics of a student studying in university $A$ and that of a student in university $B$ can be very different due to a lot of factors such as geographic location, nature of university and infrastructure. Hence the parameters for analysis of these two students should be significantly different. This paper gives the user this flexibility to configure various attributes such as weight of tags, type of membership functions and the set of recommendations. This practice would yield better results.
The long term applications are vast. This method can be used to access the performance of all students in an university. The results can be shared with the university so that the university can take appropriate actions for the welfare of the students in general.
\section*{Acknowledgment}
The authors would like to thank Professor Sudhirkumar Barai of the Civil Engineering Department, IIT Kharagpur for his continued and unconditional guidance. An exemplary teacher and a magnificent person, we consider ourselves lucky to have been taught by him and to have worked under his supervision. Without his course, \quotes{Soft Computing Tools in Engineering}, and his co-operation the preparation of this paper would not have been possible.
\vspace{34em}

\ifCLASSOPTIONcaptionsoff
  \newpage
\fi



%

\bibliography{ref} 

\begin{thebibliography}{1}

\bibitem{Patel}
S.~Patel, P.~Sajja, and A.~Patel, ``Fuzzy logic based expert system for
  students performance evaluation in data grid environment,'' {\em
  International Journal of Scientific \& Engineering Research}, vol.~5, no.~1,
  2014.

\bibitem{Chrysafiadi2015}
K.~Chrysafiadi and M.~Virvou, {\em Fuzzy Logic in Student Modeling},
  pp.~25--60.
\newblock Cham: Springer International Publishing, 2015.

\bibitem{Ingoley}
S.~Ingoley and J.~Bakal, ``Evaluation of student performance in laboratory
  applications using fuzzy logic,'' {\em International Journal on Advanced
  Computer Engineering and Communication Technologies}, vol.~1, pp.~2278 --
  5140, 2012.

\bibitem{GOKMEN2010902}
G.~Gokmen, T.~C. Akinci, M.~Tektas, N.~Onat, G.~Kocyigit, and N.~Tektas,
  ``Evaluation of student performance in laboratory applications using fuzzy
  logic,'' {\em Procedia - Social and Behavioral Sciences}, vol.~2, no.~2,
  pp.~902 -- 909, 2010.

\bibitem{Hameed}
I.~A. Hameed and C.~G. Sorensen, {\em Fuzzy Systems in Education: A More
  Reliable System for Student Evaluation}, ch.~1.
\newblock InTech, 2010.

\bibitem{Xu:2002:ISP:820741.820940}
D.~Xu, H.~Wang, and K.~Su, ``Intelligent student profiling with fuzzy models,''
  in {\em Proceedings of the 35th Annual Hawaii International Conference on
  System Sciences (HICSS'02)-Volume 3 - Volume 3}, HICSS '02, (Washington, DC,
  USA), pp.~81.2--, IEEE Computer Society, 2002.

\bibitem{Huapaya}
C.~R. Huapaya, ``Proposal of fuzzy logic-based students’learning assessment
  model,'' {\em XVIII Congreso Argentino de Ciencias de la Computacin}, 2012.

\bibitem{ZADEH1965338}
L.~Zadeh, ``Fuzzy sets,'' {\em Information and Control}, vol.~8, no.~3, pp.~338
  -- 353, 1965.

\bibitem{Klir:1996:FSF:234347}
G.~J. Klir and B.~Yuan, eds., {\em Fuzzy Sets, Fuzzy Logic, and Fuzzy Systems:
  Selected Papers by Lotfi A. Zadeh}.
\newblock River Edge, NJ, USA: World Scientific Publishing Co., Inc., 1996.

\end{thebibliography}
\bibliographystyle{ieeetr}

%




\end{document}